\documentstyle[prd,aps,multicol,epsfig]{revtex}
\newcommand{\la}{\stackrel{<}{ _{\sim}}}
\newcommand{\ga}{\stackrel{>}{ _{\sim}}}
\newcommand{\ve}[1]{\mbox{\boldmath $#1$}}
\begin{document}
\title{Dynamical instability of new-born neutron stars as sources of gravitational 
radiation}
\author{Yuk Tung Liu}
\address{Theoretical Astrophysics, California Institute of Technology,
Pasadena, California 91125}
\date{February 21, 2002}
\maketitle

\begin{abstract}
The dynamical instability of new-born neutron stars is studied by evolving 
the linearized hydrodynamical equations. The neutron stars considered in this 
paper are those produced by the accretion induced collapse of rigidly rotating 
white dwarfs. A dynamical bar-mode ($m=2$) instability is observed when the 
ratio of rotational kinetic energy to gravitational potential energy $\beta$ 
of the neutron star is greater than the critical value $\beta_d\approx 0.25$. 
This bar-mode instability leads to the emission of gravitational radiation 
that could be detected by gravitational wave detectors. However, these sources 
are unlikely to be detected by LIGO~II interferometers if the event rate is 
less than $10^{-6}$ per year per galaxy. Nevertheless, if a significant fraction 
of the pre-supernova cores are rapidly rotating, there would be a substantial number 
of neutron stars produced by the core collapse undergoing bar-mode 
instability. This would greatly increase the chance of detecting the 
gravitational radiation.
\end{abstract}



\begin{multicols}{2}

\section{Introduction}
\label{sec:intro}

Neutron stars are believed to form from the core collapse of massive stars 
and the accretion induced collapse of massive white dwarfs. If the stellar 
core or white dwarf is rotating, conservation of angular momentum implies 
that the resulting neutron star must rotate very rapidly. It has been 
suggested~\cite{thorne98} 
that such a rapidly rotating star may develop a non-axisymmetric dynamical 
instability, emitting a substantial amount of gravitational radiation which 
might be detectable by gravitational wave observatories such as LIGO, 
VIRGO, GEO and TAMA.

Rotational instabilities arise from non-axisymmetric perturbations having 
angular dependence $e^{im\varphi}$, where $\varphi$ is the azimuthal angle. 
The $m=2$ mode is called the bar mode, which is usually the strongest mode
for stars undergoing instabilities. There are two types of instabilities. 
A {\em dynamical} instability is driven by hydrodynamics and gravity, and 
it develops on a dynamical timescale, i.e.\ the timescale for a sound 
wave to travel across the star. A {\em secular} instability, on the other 
hand, is driven by viscosity or gravitational radiation reaction, and its 
growth time is determined by the relevant dissipative timescale. These secular 
timescales are usually much longer than the dynamical timescale of the system.

In this paper, we focus on the dynamical instabilities resulting from 
the new-born neutron stars formed from accretion induced collapse (AIC) of 
white dwarfs. These instabilities occur only for rapidly rotating stars. 
A useful parameter to characterize the rotation of a star is $\beta=T/|W|$, 
where $T$ and $W$ are the rotational kinetic energy and gravitational 
potential energy respectively. It is well-known that there is a critical 
value $\beta_d$ so that a star will be dynamically unstable if its 
$\beta>\beta_d$. For a uniform density and rigidly rotating star, the 
Maclaurin spheroid, the critical value is determined to be 
$\beta_d\approx 0.27$~\cite{chandrasekhar69}. Numerous numerical 
simulations using Newtonian gravity show that $\beta_d$ remains roughly 
the same for differentially rotating polytropes having the same specific 
angular momentum distribution as the Maclaurin spheroids~\cite{tohline85,durisen86,williams88,houser94,smith96,houser96,pickett96,houser98,new00}. 
However, $\beta_d$ can take values between 0.14 to 0.27 for other 
angular momentum distributions~\cite{imamura95,pickett96,centrella00}
(the lower limit $\beta_d=0.14$ is observed only for a star having 
a toroidal density distribution, i.e.\ the maximum density occurs 
off the center~\cite{centrella00}).
Numerical simulations using full general relativity and post-Newtonian 
approximations suggest that relativistic corrections to Newtonian gravity 
cause $\beta_d$ to decrease slightly~\cite{stergioulas98,shibata00,saijo00}.

Most of the stability analyses to date have been carried out by assuming 
that the star rotates with an {\it ad hoc} rotation law or using simplified 
equations of state. The results of these analyses might not be 
applicable to the new-born neutron stars resulting from AIC. Recently, 
Fryer, Holz and Hughes~\cite{fryer01} carried out an AIC simulation 
using a realistic rotation 
law and a realistic equation of state. Their pre-collapse white dwarf 
has an angular momentum $J=10^{49}~\rm{g}~\rm{cm}^2~\rm{s}^{-1}$. After 
the collapse, the neutron star has $\beta$ less than 0.06, which is 
too small for the star to be dynamically unstable. However, they point out 
that if the pre-collapse white dwarf spins faster, the resulting neutron 
star could have high enough $\beta$ to trigger a dynamical instability. 
They also point out that a pre-collapse white dwarf could easily be 
spun up to rapid rotation by accretion. The spin of an accreting white dwarf
before collapse depends on its initial mass, its magnetic field strength 
and the accretion rate, etc.~\cite{narayan89}.

Liu and Lindblom~\cite{liu01} (hereafter Paper~I) in a recent paper construct 
equilibrium 
models of new-born neutron stars resulting from AIC based on conservation 
of specific angular momentum. Their results show that if the pre-collapse white dwarfs 
are rapidly rotating, the resulting neutron stars could have 
$\beta$ as large as 0.26, which is slightly smaller than the 
critical value $\beta_d$ for 
Maclaurin spheroids. However, the specific angular momentum distributions of those
neutron stars are very different from that of Maclaurin spheroids. So
there is no reason to believe that the traditional value $\beta_d=0.27$
can be applied to those models.

The purpose of this paper is first to determine the critical value $\beta_d$ 
for the new-born neutron stars resulting from AIC, and then estimate the 
signal to noise ratio and detectability of the gravitational waves emitted 
as a result of the instability. We do not intend to provide an accurate 
number for the signal to noise ratio, which requires a detailed 
non-linear evolution of the dynamical instability. Instead, we  
use Newtonian gravitation theory to compute the structure of new-born 
neutron stars. Then we evolve the linearized Newtonian hydrodynamical 
equations to study the star's stability and determine the critical value 
$\beta_d$. Relativistic effects are expected to give a correction of 
order $(v/c)^2$, which is about 8\% for the rapidly rotating neutron stars 
studied in this paper. Here $v$ is a typical sound speed inside the star 
and $c$ is the speed of light. 

This paper is organized as follows. In Sec.~\ref{sec:eqm}, we 
apply the method described in Paper~I to construct a number 
of equilibrium neutron star models with different values of $\beta$. 
In Sec.~\ref{sec:stab}, we study the stability of these models 
by adding small density and velocity perturbations to the equilibrium 
models. Then we evolve the perturbations by solving linearized hydrodynamical 
equations proposed by Toman et al~\cite{toman98}. 
From the simulations, we can find out whether the star is stable, and 
determine the critical value $\beta_d$.
In Sec.~\ref{sec:gw}, we estimate the strength 
and signal to noise ratio of the gravitational waves emitted by this 
instability. In Sec.~\ref{sec:mag}, we estimate the effects of a magnetic 
field on the stability results. Finally, we summarize and discuss our results in 
Sec.~\ref{sec:dis}. 

\section{Equilibrium Models}
\label{sec:eqm}

In this section, we describe briefly how we construct new-born neutron
star models from the pre-collapse white dwarfs. A more detailed description 
is given in Paper~I.

\subsection{Pre-collapse white dwarf models}
\label{sec:wd}

We consider two types of pre-collapse white dwarfs: those made of carbon-oxygen 
(C-O) and those made of oxygen-neon-magnesium (O-Ne-Mg). The collapse of a massive 
C-O white dwarf is triggered by the explosive carbon burning near the center of the
star~\cite{nomoto87,nomoto91}. The central density of the pre-collapse C-O white 
dwarf must be in the range 
$6\times 10^9~\rm{g}~\rm{cm}^{-3}\la \rho_c \la 10^{10}~\rm{g}~\rm{cm}^{-3}$ 
in order for the collapse to result in a neutron star, rather than exploding 
as a Type~Ia supernova~\cite{bravo99}. The collapse of a massive O-Ne-Mg white 
dwarf, on the other hand, is triggered by electron captures by $^{24}\rm{Mg}$ and 
$^{20}\rm{Ne}$ when the central density reaches 
$4\times 10^9~\rm{g}~\rm{cm}^{-3}$~\cite{nomoto87,nomoto91}.

We construct three sequences of pre-collapse white dwarfs, with models in each 
sequence having different amounts of rotation. Sequences I and II 
correspond to C-O white dwarfs with central densities 
$\rho_c=10^{10}~\rm{g}~\rm{cm}^{-3}$ and $\rho_c=6\times 10^9~\rm{g}~\rm{cm}^{-3}$
respectively. Sequence III is for O-Ne-Mg white dwarfs with 
$\rho_c=4\times 10^9~\rm{g}~\rm{cm}^{-3}$. All white dwarfs are assumed to rotate 
rigidly, because the timescale for a magnetic field to suppress differential 
rotation is much shorter than the accretion timescale (see Sec.~\ref{sec:dis}).

The pre-collapse white dwarfs constructed in this section are described by the 
equation of state 
(EOS) of a zero-temperature ideal degenerate electron gas with electrostatic 
corrections derived by Salpeter~\cite{salpeter61}. At high density,
the pressure is dominated by the ideal 
degenerate Fermi gas with electron fraction $Z/A=0.5$ that is suitable 
for both C-O and O-Ne-Mg white dwarfs. Electrostatic corrections,
which depend on the white dwarf composition through the atomic number 
$Z$, contribute only a few percent to the EOS for the high density 
white dwarfs considered here.

Equilibrium models are computed by Hachisu's self-consistent field 
method~\cite{hachisu86}, which is an iteration scheme based on the integrated 
Euler equation for hydrostatic equilibrium:
\begin{equation}
  h+\Phi-\frac{\varpi^2}{2}\Omega^2 = C \ ,
\label{euler1}
\end{equation}
where $\Omega$ is the rotational angular frequency of the star; $C$ is a constant; 
$\varpi$ is the radius from the rotation axis;
$h$ is the specific enthalpy, which is related to the density $\rho$ and 
pressure $P$ by
\begin{equation}
  h=\int_0^P \frac{dP}{\rho}\ .
\end{equation}
The gravitational potential $\Phi$ satisfies the Poisson equation
\begin{equation}
  \nabla^2 \Phi=4\pi G \rho \ ,
\end{equation}
where $G$ is the gravitational constant. The self-consistent field method 
determines the structure of the star for fixed values of two adjustable 
parameters. In Ref.~\cite{hachisu86}, 
the maximum density and axis ratio (the ratio of polar to equatorial 
radii) are the chosen parameters. However, it is more convenient to choose 
the central density $\rho_c$ and equatorial radius $R_e$ as the two parameters
for the models studied here.

\newpage
\begin{table}
\caption{Properties of pre-collapse white dwarfs. Here $\Omega$ is the rotational 
angular frequency; $\Omega_m$ is the maximum rotational angular frequency of the 
white dwarf in the sequence without
mass-shedding; $R_e$, $R_p$, $M$, $J$ and $\beta$ are respectively the equatorial 
radius, polar radius, mass, angular momentum and the ratio of rotational kinetic 
to gravitational potential energies.}
\label{tab:eqm-wd}

\begin{center} 
Sequence I: C-O white dwarfs with $\rho_c=10^{10}~\rm{g}~\rm{cm}^{-3}$
\end{center}
\begin{tabular}{cccccc}
 $\Omega/\Omega_m$ & $R_e$ & $R_e/R_p$ & $M/M_{\odot}$ & $J$ & $\beta$ \\
 & (km) & & & ($\rm{g}~\rm{cm}^2~\rm{s}^{-1}$) & \\
\hline
  0.20 & 1310 & 1.01 & 1.40 & $5.14\times 10^{48}$ & $5.36\times 10^{-4}$ \\
  0.65 & 1400 & 1.09 & 1.42 & $1.81\times 10^{49}$ & $6.08\times 10^{-3}$ \\
  0.84 & 1500 & 1.19 & 1.44 & $2.43\times 10^{49}$ & $1.03\times 10^{-2}$ \\
  0.93 & 1600 & 1.27 & 1.46 & $2.80\times 10^{49}$ & $1.30\times 10^{-2}$ \\
  1.00 & 1895 & 1.52 & 1.47 & $3.12\times 10^{49}$ & $1.55\times 10^{-2}$ \\
\end{tabular}
\begin{center} 
Sequence II: C-O white dwarfs with $\rho_c=6\times 10^{10}~\rm{g}~\rm{cm}^{-3}$
\end{center}
\begin{tabular}{cccccc}
 $\Omega/\Omega_m$ & $R_e$ & $R_e/R_p$ & $M/M_{\odot}$ & $J$ & $\beta$ \\
 & (km) & & & ($\rm{g}~\rm{cm}^2~\rm{s}^{-1}$) & \\
\hline
 0.23 & 1517 & 1.01 & 1.39 & $3.31\times 10^{48}$ & $7.73\times 10^{-4}$ \\
 0.64 & 1610 & 1.09 & 1.42 & $1.95\times 10^{49}$ & $6.11\times 10^{-3}$ \\
 0.84 & 1740 & 1.19 & 1.44 & $2.75\times 10^{49}$ & $1.12\times 10^{-2}$ \\
 0.93 & 1847 & 1.28 & 1.45 & $3.13\times 10^{49}$ & $1.39\times 10^{-2}$ \\
 1.00 & 2189 & 1.52 & 1.46 & $3.51\times 10^{49}$ & $1.66\times 10^{-2}$ \\
\end{tabular}
\begin{center}
Sequence III: O-Ne-Mg white dwarfs with $\rho_c=4\times 10^{10}~\rm{g}~\rm{cm}^{-3}$
\end{center}
\begin{tabular}{cccccc}
 $\Omega/\Omega_m$ & $R_e$ & $R_e/R_p$ & $M/M_{\odot}$ & $J$ & $\beta$ \\
 & (km) & & & ($\rm{g}~\rm{cm}^2~\rm{s}^{-1}$) & \\
\hline
 0.23 & 1692 & 1.01 & 1.38 & $7.01\times 10^{48}$ & $7.90\times 10^{-4}$ \\
 0.62 & 1791 & 1.09 & 1.40 & $2.05\times 10^{49}$ & $6.23\times 10^{-3}$ \\
 0.86 & 1956 & 1.20 & 1.42 & $3.03\times 10^{49}$ & $1.23\times 10^{-2}$ \\
 0.96 & 2156 & 1.34 & 1.44 & $3.59\times 10^{49}$ & $1.62\times 10^{-2}$ \\
 1.00 & 2441 & 1.52 & 1.45 & $3.80\times 10^{49}$ & $1.77\times 10^{-2}$ \\
\end{tabular}
\end{table}

The accuracy of the equilibrium models can be measured by the quantity
\begin{equation}
\epsilon=\left| \frac{2T+W+3\Pi}{W}\right| \ ,
\label{def:epsilon}
\end{equation}
which should be equal to zero according to the Virial theorem (see 
e.g.~\cite{tassoul78}). Here
$T$ is the rotational kinetic energy; $W$ is the gravitational
potential energy and $\Pi=\int P d^3x$. The values of
$\epsilon$ are of order $10^{-7}$ for all the pre-collapse white
dwarf models calculated in this section.
Table~\ref{tab:eqm-wd} shows some properties of several pre-collapse white dwarfs 
in the three sequences. Each sequence terminates when the rotational angular 
frequency $\Omega$ of the white dwarf reaches a critical value $\Omega_m$ so that the 
mass-shedding occurs on the equatorial surface of the star. The values 
of $\Omega_m$ are $5.37~\rm{rad}~\rm{s}^{-1}$ for Sequence I, 
$4.32~\rm{rad}~\rm{s}^{-1}$ for Sequence II, and 
$3.65~\rm{rad}~\rm{s}^{-1}$ for Sequence III.

\begin{figure}
\epsfig{file=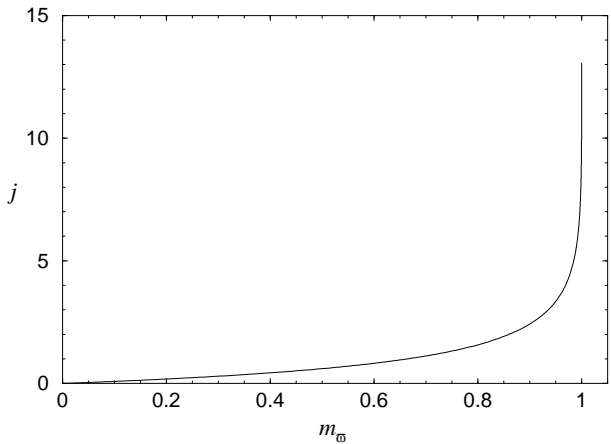,width=6cm,angle=270}
\caption{The normalized specific angular momentum $j$ as a function of the cylindrical 
mass fraction $m_{\varpi}$ for the white dwarf model in sequence III with 
$\Omega/\Omega_m=0.964$.}
\label{fig:jm}
\end{figure}

Fig.~\ref{fig:jm} shows the normalized specific angular momentum
\begin{equation}
  j=\frac{M}{J} \Omega \varpi^2
\label{eq:j}
\end{equation}
as a function of the cylindrical mass fraction 
\begin{equation}
  m_{\varpi}=\frac{2\pi}{M}\int_0^{\varpi}d\varpi'\, \varpi'\int_{-\infty}^{\infty} 
dz'\, \rho(\varpi',z')
\end{equation}
for a typical pre-collapse
model. Here $M$ and $J$ are the total mass and angular momentum of the star 
respectively. The specific angular momentum defined in Eq.~(\ref{eq:j}) is 
normalized so that $\int_0^1 j(m_{\varpi})\, dm_{\varpi} =1$.
The curves for other models are similar. The spike near $m_{\varpi}=1$ 
is due to the high degree of central condensation of the white dwarf 
density (Paper~I).

\subsection{Collapsed objects}
\label{sec:ns}

The gravitational collapse of a massive white dwarf is halted when the 
core density reaches nuclear density. The core bounces back and 
settles down into hydrodynamical equilibrium in a few milliseconds. A 
hot ($T\ga 20~\rm{Mev}$) and lepton-rich {\em proto-neutron star} is formed.
After about 20~s, neutrinos carry away most of the energy and the star cools 
down to a cold neutron star. The hot proto-neutron stars are 
less compact and have $\beta$ smaller than 0.14 (Paper~I). They 
are thus expected to be dynamically stable. Hence in this paper, we focus on  
the stability of the cold neutron stars shortly after the cooling.

We assume that 
(1) the neutron stars are axisymmetric and are in rotational equilibrium 
with no meridional circulation; (2) viscosity can be neglected; (3) no 
material is ejected during the collapse. Under these assumptions, it 
is easy to prove that (see e.g.~\cite{liu01,tassoul78}) the specific 
angular momentum $j$ of the collapsed star as a function of cylindrical 
mass fraction $m_{\varpi}$ is the same as the pre-collapse white dwarf.
Hence the structure of the new-born neutron stars can be 
constructed by computing models with the same masses, angular momentum 
and $j(m_\varpi)$-distributions as the pre-collapse white dwarfs.

We adopt the Bethe-Johnson EOS~\cite{bethe74} for densities above 
$10^{14}~\rm{g}~\rm{cm}^{-3}$, and BBP EOS~\cite{baym71} for densities 
in the range $10^{11}-10^{14}~\rm{g}~\rm{cm}^{-3}$. The EOS for densities 
below $10^{11}~\rm{g}~\rm{cm}^{-3}$ is joined by that of the pre-collapse 
white dwarfs. 

We construct the equilibrium models by the numerical method proposed by
Smith and Centrella~\cite{smith92}, which is a modified version of Hachisu's 
self-consistent field method so that $j(m_{\varpi})$ can be specified. 
The iteration scheme is based on the integrated Euler equation~(\ref{euler1}) 
written in the form
\begin{equation}
h=C-\Phi+\left(\frac{J}{M}\right)^2 \int_0^{\varpi} d\varpi'\, 
\frac{j^2(m_{\varpi'})}{\varpi'^3} \ ,
\end{equation}
where $J$ and $M$ are the total angular momentum and mass of the star 
respectively. As before, two parameters have to be fixed in the iteration 
procedure. We choose to fix the central density $\rho_c$ and equatorial 
radius $R_e$. Since the correct $\rho_c$ and $R_e$ are not known beforehand, 
we have to vary these two quantities until the equilibrium model has the 
same $J$ and $M$ as the pre-collapse white dwarf.

The standard iteration algorithm described in Refs.~\cite{hachisu86} and~\cite{smith92} 
fails to converge when the star becomes very flattened. This problem is fixed 
by a modified scheme proposed by Pickett, Durisen and Davis~\cite{pickett96}, 
in which only a fraction of the revised density (or enthalpy) $\rho_{i+1}$, i.e.\ 
$\rho'_{i+1}=(1-\zeta) \rho_{i+1}+\zeta \rho_i$, is used for the next 
iteration. Here $\zeta<1$ is a parameter controlling the change of 
density. A value of $\zeta > 0.95$ has to be used for very flattened 
configurations, and it takes 100-200 iterations for the density and 
enthalpy distributions to converge.

Paper~I mentions another numerical difficulty which has to do 
with the spike of the $j(m_{\varpi})$ curve near $m_{\varpi}=1$.
The authors of Paper~I have to truncate a small 
portion of the $j(m_{\varpi})$ curve in order to make the iteration converge. 
They also demonstrate that this truncation does not affect the inner 
structure of the star. It turns out that, for reasons still to be understood, the 
numerical instability associated with the $j(m_{\varpi})$ curve only 
occurs for the most rapidly rotating models (i.e.\ those models where 
the pre-collapse white dwarfs have $\Omega/\Omega_m=1$). Hence, the 
truncation is not necessary for all the other cases.

\begin{figure}
\epsfig{file=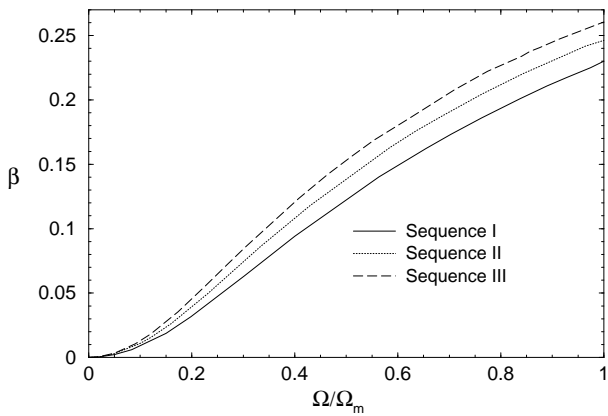,width=5.5cm,angle=270}
\caption{The values of $\beta$ of the resulting neutron stars as a function 
of $\Omega/\Omega_m$ of the pre-collapse white dwarfs.}
\label{fig:ToW}
\end{figure}

As in the case of pre-collapse white dwarfs, we measure the accuracy of the 
equilibrium models by the quantity $\epsilon$ defined in Eq.~(\ref{def:epsilon}).
The models computed in this subsection have $\epsilon$ ranges from about
$10^{-6}$ (for slowly rotating stars) to $10^{-4}$ (for rapidly rotating stars).
We construct a number of neutron star models resulting from the collapse of 
the three sequences of pre-collapse white dwarfs in the previous subsection. 
The ratio of rotational kinetic energy to gravitational potential energy, $\beta$, 
of the neutron stars is plotted in Fig.~\ref{fig:ToW} as a function 
of $\Omega/\Omega_m$ of the pre-collapse white dwarf. The values of $\beta$ 
for all the neutron star models are smaller than 0.27, the critical value of 
$\beta$ for the dynamical instability of rigidly rotating Maclaurin spheroids. 

\begin{figure}
\epsfig{file=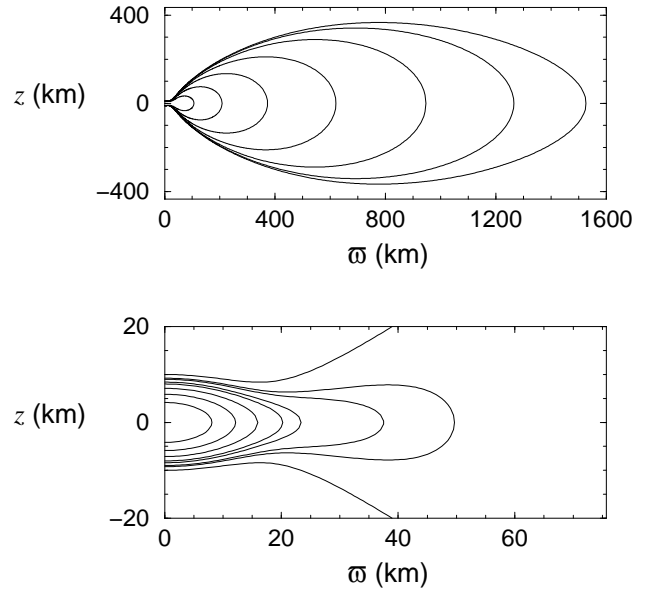,width=8cm,angle=270}
\caption{Meridional density contours of the neutron star resulting from 
the AIC of a rigidly rotating O-Ne-Mg white dwarf 
with $\Omega/\Omega_m=0.964$. This neutron star has $\beta=0.255$. The 
contours in the upper graph denote, from inward to outward, $\rho/\rho_c=
10^{-4},\ 10^{-5},\ 10^{-6},\ 10^{-7},\ 10^{-8},\ 10^{-9}\ \rm{and}\ 0$. 
The contours in the lower graph denote, from inward to outward, $\rho/\rho_c=
0.8,\ 0.6,\ 0.4,\ 0.2,\ 0.1,\ 10^{-2},\ 10^{-3}\ \rm{and}\ 10^{-4}$. The 
central density of the star is $\rho_c=3.3\times 10^{14}~\rm{g}~\rm{cm}^{-3}$.}
\label{dencont}
\end{figure}

The structure of the neutron stars with $\beta \ga 0.1$ are all similar: 
they contain a high-density 
central core of size about 20~km, surrounded by a low-density torus-like envelope.
The size of the envelope depends on the amount of rotation of the star, which 
can be measured by $\beta$. The size ranges from 100~km (for $\beta \sim 0.1$) to 
over 500~km (for $\beta \ga 0.2$). Figure~\ref{dencont} shows the density 
contours of a typical model. This model corresponds to the collapse of an
O-Ne-Mg white dwarf with $\Omega/\Omega_m=0.964$. The resulting neutron star 
has $\beta=0.255$. The envelope extends to about 1530~km in this case. As a 
comparison, the equatorial radius of the pre-collapse white dwarf is 2156~km.

\begin{figure}
\epsfig{file=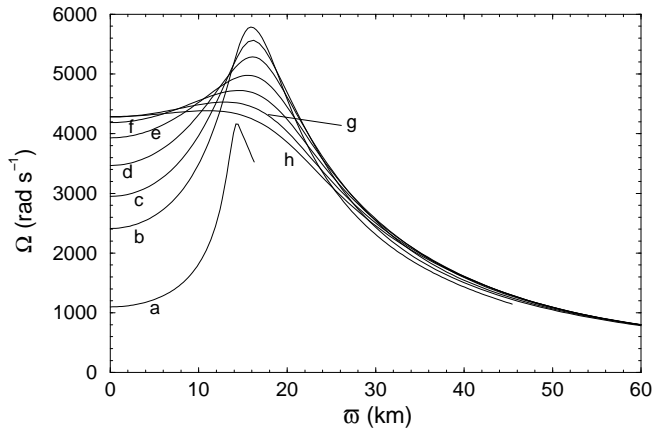,width=5.8cm,angle=270}
\caption{The distribution of rotational angular velocity $\Omega$ as a function 
of $\varpi$ for $\varpi<60~\rm{km}$. These are models for sequence III with 
$\beta$'s of the resulting neutron stars equal to (a) 0.0106, (b) 0.0555, (c) 0.0860, 
(d) 0.124, (e) 0.169, (f) 0.208, (g) 0.238 and (h) 0.261. The equatorial radii 
of the neutron stars in cases (a) and (b) are smaller than 60~km, and their 
frequency curves terminate at their equatorial radii.}
\label{fig:om-III}
\end{figure}

\begin{figure}
\epsfig{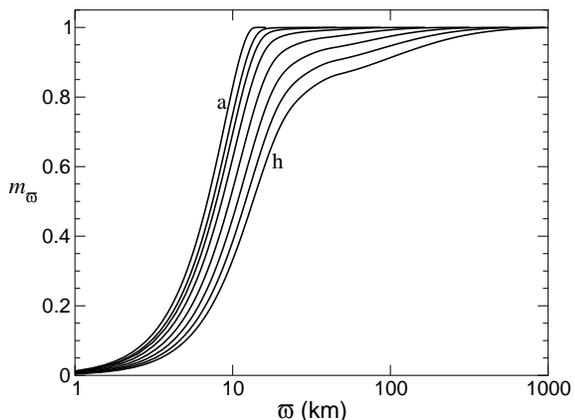}
\caption{The cylindrical mass fraction $m_{\varpi}$ as a function of $\varpi$ for 
neutron star models in Fig.~\ref{fig:om-III}. The curves and their corresponding 
models are identified by the degree of central condensation: the higher the 
degree of central condensation, the lower the value of $\beta$.}
\label{fig:mp-III}
\end{figure}

\begin{figure}
\epsfig{file=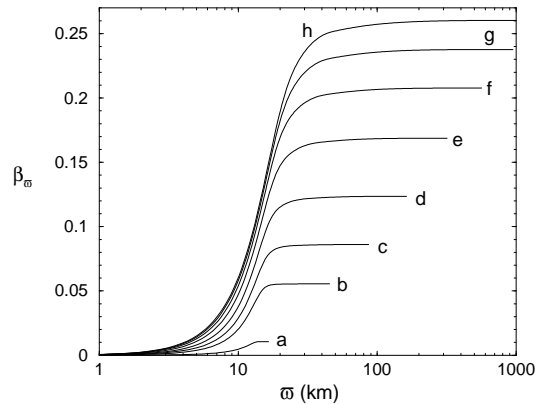,width=5.4cm,angle=270}
\caption{The value of $\beta_{\varpi}$ as a function of $\varpi$ for the 
neutron star models in Fig.~\ref{fig:om-III}. The curves for models (a)-(g) 
terminate at the equatorial radii of the stars.}
\label{fig:ToverW-III}
\end{figure}

Figure~\ref{fig:om-III} shows the rotational angular velocity $\Omega$ as 
a function of radius $\varpi$ for neutron star models corresponding to the 
collapse of sequence III white dwarfs. The cases for sequences~I and II 
are similar. We see that stars with small $\beta$ show strong 
differential rotation. However, the rotation in the core region ($\varpi \la 
20~\rm{km}$) becomes more and more rigidly rotating as $\beta$ increases. 
The most rapidly rotating case ($\beta=0.261$) has $\Omega \approx 
4400~\rm{rad}~\rm{cm}^{-1}$ in the core. This corresponds to a rotation 
period of 1.4~ms, slightly less than the period of the fastest observed 
millisecond pulsar (1.56~ms). Further analysis reveals that the rotation curve
in the envelope region roughly follows the Kepler law 
$\Omega \propto \varpi^{-3/2}$.

Figure~\ref{fig:mp-III} shows the cylindrical mass fraction $m_{\varpi}$ as 
a function of $\varpi$ for the same models as in Fig.~\ref{fig:om-III}. As expected, 
the degree of central concentration decreases with increasing $\beta$. However, 
more than 80\% of the mass is still concentrated inside a radius $\varpi=30~\rm{km}$, 
even for the most rapidly rotating case. The collapsed object can be regarded 
as a rotating neutron star surrounded by an accretion torus.

Numerous numerical studies demonstrate that the quantity $\beta$ is an 
important parameter for the dynamical stability of a rotating star. It is 
then useful to define a function $\beta_{\varpi}$ as 
\begin{equation}
  \beta_{\varpi}=\frac{\int_0^{\varpi} d\varpi'\, \varpi' \int_{-\infty}^{\infty}
dz'\, [\varpi' \Omega(\omega')]^2 \rho(\varpi',z')}
{ |\int_0^{\varpi} d\varpi'\, \varpi' \int_{-\infty}^{\infty} dz'\, 
\rho(\varpi',z') \Phi(\varpi',z')|} \ ,
\label{def:betapm}
\end{equation}
which is a measure of $\beta$ for the material inside a cylinder of radius 
$\varpi$ from the rotation axis. Figure~\ref{fig:ToverW-III} plots $\beta_{\varpi}$ 
as a function of $\varpi$ for the same neutron star models as in Fig.~\ref{fig:om-III}.
We see that the curves level off when $\varpi \ga 20-100~\rm{km}$ for all 
rapidly rotating models, suggesting that the material outside 100~km is 
probably unimportant for dynamical stability. It should also be noticed that 
the major contribution to $\beta$ is from the region $10~\rm{km} \la \varpi \la 
50~\rm{km}$. Hence we expect that the material in this region plays an 
important role on the dynamical stability of the star.

\section{Stability of the Collapsed Objects}
\label{sec:stab}

In this Section, we study the dynamical stabilities of the neutron star models 
computed in Sec.~\ref{sec:ns} using the technique of the linear stability 
analysis developed by Toman et al~\cite{toman98}. This technique is briefly 
reviewed in Sec.~\ref{sec:LSA}. We then  report the stability results in 
Sec.~\ref{sec:LSAresult}. 

\subsection{Linear stability analysis}
\label{sec:LSA}

The motion of fluid inside the star is described by the hydrodynamical equations:
\begin{eqnarray}
  \partial_t \rho + \nabla_a (\rho v^a) &=& 0 \label{eq:con} \ , \\
  \partial_t v^a+v^b \nabla_b v^a &=& -\frac{\nabla^a P}{\rho} - \nabla^a \Phi
\label{eq:euler} \ , \\
  \nabla_a \nabla^a \Phi = 4\pi G \rho \ ,
\label{eq:poisson}
\end{eqnarray}
where the summation convention is assumed, and $\nabla_a$ denotes the covariant 
derivative compatible with three-dimensional flat-space metric.
To study the stability, we perturb the density $\rho$ and velocity $v^a$ 
away from their equilibrium values by small quantities:
\begin{eqnarray}
  \rho(x^b,t) &=& \rho_0(x^b)+\delta \rho(x^b,t) \ , \\
  v^a(x^b,t) &=& \varpi \Omega(\varpi) e^a_{\hat{\varphi}}+
\delta v^a(x^b,t) \ ,
\end{eqnarray}
where $e^a_{\hat{\varphi}}$ is the unit vector along the azimuthal direction.
The Lagrangian pressure perturbation $\Delta P$ is related to the Lagrangian 
density perturbation $\Delta \rho$ by 
\begin{equation}
  \Delta P = \gamma_p \frac{P}{\rho} \Delta \rho \ ,
\end{equation}
where for simplicity, the subscript ``0'' is suppressed, and hereafter in 
this Section, $\rho$ and $P$ denote the equilibrium density and pressure 
respectively. The quantity 
\begin{equation}
  \gamma_p = \left( \frac{d \log P}{d \log \rho} \right)_p
\end{equation}
is the adiabatic index for pulsation. The relation between the Eulerian 
perturbations $\delta P$ and $\delta \rho$ can be easily deduced from the 
transformation between the Lagrangian and Eulerian perturbations. The 
result is 
\begin{equation}
  \delta P = \gamma_p \frac{P}{\rho} \delta \rho + \left( \frac{\gamma_p}
{\gamma_{\rm eq}} -1\right) \xi^a \nabla_a P \ ,
\end{equation}
where 
\begin{equation}
  \gamma_{\rm eq} = \left( \frac{d \log P}{d \log \rho} \right)_{\rm eq}
\end{equation}
is the adiabatic index computed from the equilibrium EOS. The Langrangian 
displacement $\xi^a$ satisfies the equation 
\begin{equation}
  \partial_t \xi^a + v^b \nabla_b \xi^a -\xi^b \nabla_b v^a 
= \delta v^a \ .
\end{equation}
The Eulerian change of the gravitational potential $\delta \Phi$ satisfies 
the Poisson equation 
\begin{equation}
 \nabla_a \nabla^a \delta \Phi = 4\pi G \delta \rho \ .
\end{equation}
We find it useful to introduce a quantity $\delta h \equiv \delta P/\rho$, 
which is related to $\delta \rho$ by 
\begin{equation}
  \delta h = \gamma_p \frac{P}{\rho^2} \delta \rho + \left( \frac{\gamma_p}
{\gamma_{\rm eq}} -1\right) \xi^a \nabla_a h \ .
\end{equation}
In the region where $\gamma_p=\gamma_{\rm eq}$, $\delta h$ is the Eulerian 
change of the enthalpy.

If the system is unstable, the perturbed quantities will grow in time. Instead 
of solving the fully non-linear equations~(\ref{eq:con})--(\ref{eq:poisson}), 
however, Toman et al~\cite{toman98} develop a more efficient approach: expand  
Eqs.~(\ref{eq:con})--(\ref{eq:poisson}) to linear order of the perturbations 
and evolve the linearized equations.

Consider the angular Fourier decomposition of any perturbed quantity $\delta q$:
\begin{equation}
  \delta q(x^b,t)=\sum_{m=-\infty}^{\infty} \delta \tilde{q}_m(\varpi,z,t) 
e^{im\varphi} \ .
\end{equation}
It can be easily proved that each $m$-mode decouples in the linearized hydrodynamical 
equations because of the axisymmetry of the equilibrium configuration. In addition, 
the fact that the equilibrium configuration is symmetric under reflection 
about the equatorial plane ($z\rightarrow -z$) implies the modes with even 
and odd parity under the transformation $z\rightarrow -z$ also decouple.
Hence each $m$-mode with a definite parity 
can be evolved separately and the 3+1 simulation is reduced to a 2+1 simulation, 
which saves a lot of computation time. Hereafter, all perturbed quantities 
will be assumed to have angular dependence $e^{im\varphi}$.

In Ref.~\cite{toman98}, Toman et al choose to evolve the variables $\delta \rho$ 
and $\delta v^a$. However, we find it more convenient and numerically stable 
in our case to evolve 
the variables $\delta h$ and $\delta p^a = \rho \delta v^a$. The reason being 
that the simulations are performed on a discrete grid, and it is preferable 
to use variables that change smoothly to ensure accuracy. However, 
the background density $\rho$ decreases abruptly outside the core region, 
and the perturbation $\delta \rho$ is expected to behave similarly. On the 
other hand, $h$, and presumably $\delta h$, change much more smoothly even near the 
boundary of the star. In the case where $\gamma_p \neq \gamma_{\rm eq}$, 
we also need to evolve the scalar function $\eta = \xi^b \nabla_b h$. 
In terms of the new variables, the linearized equations become
\begin{eqnarray}
  \partial_t \delta h &=& -im\Omega \delta h -\gamma_p \frac{P}{\rho^2}\nabla_a 
\delta p^a \cr
 & & + \left(\frac{\gamma_p}{\gamma_{\rm eq}}-1\right)\frac{\delta p^a}{\rho}
\nabla_a h \label{eq:lcon} \ , \\
  \partial_t \delta p^a &=& -\delta p^b\nabla_b v^a 
-v^b\nabla_b \delta p^a - \rho \nabla^a \delta h -\rho \nabla^a \delta \Phi  \cr
 & & -\left(\frac{\gamma_p}{\gamma_{\rm eq}}-1\right) 
\frac{\rho^2}{\gamma_p P}(\delta h +\eta) \nabla^a h \label{eq:leuler} \ , \\
 \partial_t \eta &=& -im\Omega \eta +\frac{\delta p^a}{\rho} \nabla_a h \ , \\
  \nabla_a \nabla^a \delta \Phi &=& \frac{4\pi G \rho^2}{\gamma_p P}\left[ 
\delta h -\left(\frac{\gamma_p}{\gamma_{\rm eq}}-1\right)\eta\right] \ .
\label{eq:lpoisson}
\end{eqnarray}
It follows from Eqs.~(\ref{eq:lcon})--(\ref{eq:lpoisson}) that 
if $(\delta h,\delta p^a)$ 
is a solution for an $m$-mode, the complex conjugate $(\delta h^*,{\delta p^a}^*)$ 
is a solution for the -$m$-mode. We can then define the physical ``enthalpy''
perturbation $\delta \tilde{h}=\delta h + \delta h^*$, and similarly for the 
physical density $\delta \tilde{\rho}$ and velocity $\delta \tilde{v}^a$ 
perturbations of an $m$-mode.

We use a uniform cylindrical grid to perform the simulations. We have checked that the 
code is able to reproduce the results in Ref.~\cite{toman98}. However, unlike 
the case in Ref.~\cite{toman98}, the collapsed objects studied here 
have a large envelope extending beyond 1000~km when the stars under consideration 
are rapidly rotating. This numerical difficulty can be circumvented by 
a suitable truncation scheme.

As pointed out in Sec.~\ref{sec:ns}, we expect the outer envelope will not 
influence the dynamical stability in any significant way.
Hence it is necessary to evolve the perturbations 
only in the dynamically interesting region. This is done by introducing a 
radius $R_m$ and a minimum density $\rho_{\rm{min}}\equiv \rho(R_m,0)$. 
The perturbations are set to zero wherever the equilibrium 
density $\rho(\varpi,z)<\rho_{\rm{min}}$. If $R_m$ is sufficiently large, 
increasing its value will not change the evolution result. We find that a value 
of $R_m\approx 200~\rm{km}$ is needed to ensure that the results converge, 
and we use a cylindrical grid with $400\times 400$ grid points to achieve a 
resolution of 0.5~km.

In general, the two adiabatic indices $\gamma_p$ and $\gamma_{\rm{eq}}$ are 
not equal.
They coincide only if the pulsation timescale is much longer than all 
the reaction timescales for the different species of particles in the fluid to achieve 
equilibrium. This is the case for densities below neutron drip ($\rho \la 
4\times 10^{11}~\rm{g}~\rm{cm}^{-3}$) and above about $10^{13}~\rm{g}~\rm{cm}^{-3}$.
However, in the density range $4\times 10^{11}~\rm{g}~\rm{cm}^{-3} \la \rho \la 
10^{13}~\rm{g}~\rm{cm}^{-3}$, the matter is a mixture of electrons, neutrons 
and nuclei in equilibrium. Some of the reactions required to achieve equilibrium 
involve weak interactions, which have timescales much longer than the pulsation 
timescale. Hence equilibrium is not achieved during pulsation, and $\gamma_p 
\neq \gamma_{\rm{eq}}$ in that density range~\cite{meltzer66,colpi89}. Most people 
studying neutron star 
pulsations neglect the difference of $\gamma_p$ and $\gamma_{\rm{eq}}$ and 
use $\gamma_{\rm{eq}}$ in their calculations. It has been demonstrated 
(see e.g.~\cite{lindblom83}) that this treatment has no significant effect
on the final result, because the matter in that density range occupies 
only a tiny fraction of neutron star. However, it may have an important 
effect on the stability of the new-born neutron stars studied here. 
The reason is that the dynamically important region, as pointed out in 
Sec.~\ref{sec:ns}, is $10~\rm{km} \la \varpi \la 50~\rm{km}$. This region 
contains a significant amount of matter in that density range 
(see Fig.~\ref{dencont}). Our numerical simulations indicate that this is 
indeed the case. The critical value $\beta_d$ for the dynamical instability 
drops from about 0.25 to 0.23 if $\gamma_{\rm{eq}}$ is used for the adiabatic 
index of pulsation.

\begin{figure}
\epsfig{file=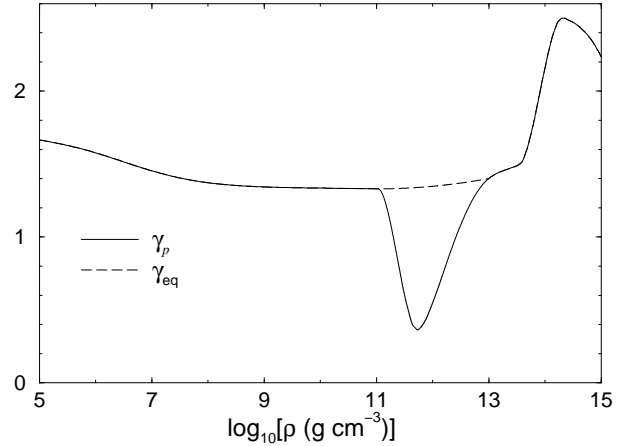,width=6cm,angle=270}
\caption{The values of $\gamma_{\rm eq}$ (solid line) and $\gamma_p$ (dashed 
line) as a function of $\log_{10} \rho$. The two curves coincide when 
$\rho<\rho_d$ and $\rho > \rho_e$.}
\label{fig:gamma}
\end{figure}

The appropriate $\gamma_p$ remains roughly constant from the density 
of neutron drip $\rho_d=4\times 10^{11}~\rm{g}~\rm{cm}^{-3}$ to the density 
above which $\gamma_{\rm{eq}}=\gamma_p$
around $\rho_e=10^{13}~\rm{g}~\rm{cm}^{-3}$~\cite{meltzer66,colpi89}. To 
provide a reasonable value of $\gamma_p$ which mimics the curve in 
Refs.~\cite{meltzer66,colpi89} and which is 
compatible with the EOS used here, we take $\gamma_p$ in the density 
range $\rho_d<\rho<\rho_e$ to be (also see
Fig.~\ref{fig:gamma})
\begin{equation}
  \gamma_p (\rho) = \gamma_{\rm{eq}}(\rho_d)+\left[ \gamma_{\rm{eq}}(\rho_e) 
-\gamma_{\rm{eq}}(\rho_d)\right] \frac{\log^2 (\rho/\rho_d)}
{\log^2 (\rho_e/\rho_d)} \ .
\label{eq:gammap}
\end{equation}

Under some circumstances it is possible to have a region of the star 
where the mode is stationary in the fluid's co-rotating frame. 
In this case, we should use $\gamma_{\rm{eq}}$
for the adiabatic index of pulsation in the region where $|\omega'|=
|\omega+m\Omega|
\ll 2\pi/t_r$. Here $\omega$ is the angular frequency of an $m$-mode 
that has dependence $\exp[i(\omega t+m\varphi)]$ in the inertial frame;
$\omega'=\omega+m\Omega$ is the angular frequency of the mode 
in the fluid's co-rotating frame;
and $t_r \approx 1~\rm{s}$ is the timescale for different species
of particles to achieve $\beta$-equilibrium in the density range
$\rho_d<\rho<\rho_e$. It turns out (see the next subsection) that 
rapidly rotating neutron stars have an unstable bar mode ($m=2$) 
with $\omega \approx -3000~\rm{rad}~\rm{s}^{-1}$. There is indeed 
a radius at which $\omega'=0$. This radius is at $\varpi=\varpi_c 
\approx 40~\rm{km}$ for stars with $\beta >0.23$. 
The density on the equator of the stars is $\rho \approx 
10^{12}~\rm{g}~\rm{cm}^{-3}$,
well within the questionable density range. However, the width 
of this ``co-rotating region'' which satisfies $|\omega'|<2\pi/t_r$ is
\[ \Delta \varpi = \frac{2\pi/t_r}{\left| \partial_{\varpi} \Omega(\varpi_c)
\right|}\approx 0.1~\rm{km} \ . \]
The material in the region contains only $10^{-4}$ of total mass and
angular momentum of the star. Hence this thin co-rotating layer is not 
expected to have a significant influence on the overall stability of
the stars.

\subsection{Results}
\label{sec:LSAresult}

We perform a number of simulations on neutron star models 
computed by the method described in Sec.~\ref{sec:ns}. The simulations 
are terminated either when an instability is fully developed or when the 
simulation time reaches 60~ms, corresponding to 40 rotation periods 
of the material at the center of the star. We regard a star as dynamically 
unstable if the density perturbation shows an evidence of exponential growth and 
increases its amplitude by at least a factor of fifteen by the end of 
the simulation. In our simulations, no 
instability is observed for neutron star models in sequences I and II. 
A bar-mode ($m=2$) instability develops for sequence III models when the
star's $\beta$ is greater than a critical value $\beta_d \approx 0.25$. 
The unstable mode has even parity under reflection about the equatorial 
plane. This $\beta_d$ is slightly less than the critical value 0.27 for the 
Maclaurin spheroids. It should be pointed out that all the stars in sequences 
I and II have $\beta$'s smaller than this $\beta_d$. Hence we believe 
that they are stable simply because their $\beta$'s are not high enough. 

Some other simulations~\cite{pickett96,centrella00} show that in the cases 
where $\beta_d<0.27$, the instability is dominated by the $m=1$ mode for 
stars with $\beta$ close to $\beta_d$. However, we do not observe any 
sign of an unstable $m=1$ mode in our case. We also performed simulations 
using $\gamma_{\rm eq}$ (the solid curve in Fig.~\ref{fig:gamma}) as 
the adiabatic index for pulsation instead of $\gamma_p$
(the dashed curve in Fig.~\ref{fig:gamma}). We find that $\beta_d$ 
drops to about 0.23, showing that matter in the density region 
$\rho_d<\rho<\rho_e$ plays an important role on the instability. 
\vskip 0.5cm
\begin{figure}
\epsfig{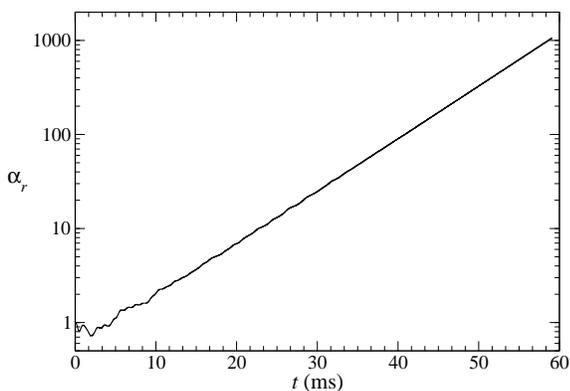}
\caption{The relative amplitude $\alpha_r$ 
as a function of time. The equilibrium star has $\beta=0.261$, the most rapidly 
rotating model.}
\label{fig:alpha}
\end{figure}

To visualize the instability, we define an amplitude 
\begin{equation}
  \alpha = \left( \frac{\int |\delta \rho|^2\, d^3 x}
{\int \rho^2\, d^3 x} \right)^{1/2} 
\label{def:alpha}
\end{equation}
for the density perturbation.
Since we evolve the perturbations using linearized equations, it is more 
convenient to work with the relative amplitude $\alpha_r$:
\begin{equation}
  \alpha_r(t) = \alpha(t)/\alpha(0) \ .
\label{def:alphar}
\end{equation}
This relative amplitude is defined so that it is equal to one at $t=0$. 
Figure~\ref{fig:alpha} shows the time evolution of $\alpha_r$ for the most rapidly 
rotating star ($\beta=0.261$). We see that after about 10~ms, an instability 
develops and $\alpha_r$ grows exponentially. The e-folding time of the growth 
$\tau$ is found, by least square fit, to be 7.8~ms. 
\vskip 0.55cm
\begin{figure}
\epsfig{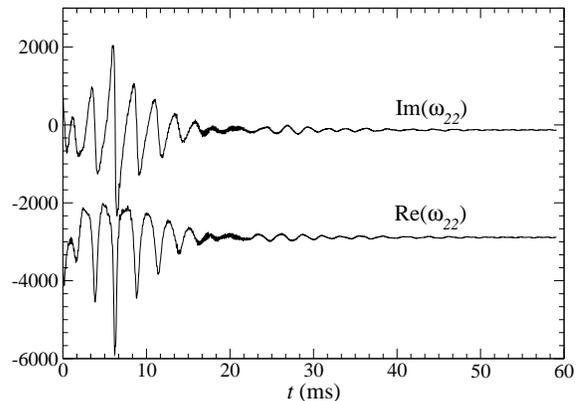}
\caption{The time evolution of the angular frequency $\omega_{22}$ for the most 
rapidly rotating star ($\beta=0.261$).}
\label{fig:om22}
\end{figure}

The unstable mode can also be characterized by a complex angular frequency defined as 
\begin{equation}
  \omega_{22}=\frac{\dot{D}_{22}}{i D_{22}} \ ,
\label{def:om22}
\end{equation}
where 
\begin{equation}
  D_{22}=\int r^2 \delta \rho Y^*_{22}\, d^3 x
\label{def:d22}
\end{equation}
is the mass quadrupole moment, and the spherical harmonic function
\[ 
Y_{22} = \frac{1}{4} \sqrt{\frac{15}{2\pi}}\, \sin^2 \theta \, e^{2i\varphi} \ . 
\]
The time derivative of $D_{22}$ is evaluated by the formula~\cite{finn90}
\begin{equation}
  \dot{D}_{22} = \int \delta p^a \nabla_a (r^2 Y^*_{22})\, d^3 x - 
2i \int \Omega r^2 \delta \rho Y^*_{22}\, d^3 x \ , 
\end{equation}
where we have used the continuity equation~(\ref{eq:con}) and integrated  
by parts.

Let $\omega$ be the complex frequency of the most unstable mode. The e-folding 
time is related to the imaginary part of $\omega$ by $\rm{Im}(\omega)=-1/\tau$.
At late time, the density perturbation is dominated by the most unstable mode,
which means that both $\delta \rho$ and $D_{22}$ go approximately as $\exp(i\omega t)$. 
Hence $\omega_{22}\approx \omega$. Fig.~\ref{fig:om22} plots $\omega_{22}$ as a 
function of time for the evolution of the most rapidly rotating star. We see 
that at late time, $\omega_{22}$ is approximately a constant,
indicating that the perturbation is indeed dominated by the most 
unstable mode. 
The frequency of the unstable mode is then determined to be $\omega \approx 
(-2890-130i)~\rm{rad}~\rm{s}^{-1}$. Note that the imaginary part agrees with 
the e-folding time determined above.
\vskip 0.6cm
\begin{figure}
\epsfig{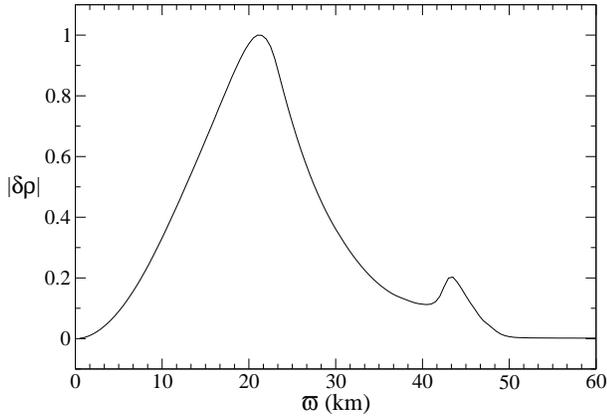}
\caption{The magnitude of the density perturbation $|\delta \rho|$ 
of the unstable bar mode of the most rapidly rotating star ($\beta=0.261$) 
on the equatorial plane. The magnitude is normalized so that the maximum 
value is one.}
\label{fig:cdeneq}
\end{figure}

Figure~\ref{fig:cdeneq} shows the magnitude of the density perturbation 
$|\delta \rho|$ of the unstable bar mode of the most rapidly rotating star 
($\beta=0.261$) on the equatorial plane. We see that $|\delta \rho|$ has 
a peak at $\varpi \approx 20~{\rm km}$, which is in the transition region between 
the neutron star core and the tenuous outer layers (see Fig.~\ref{dencont}). 
There is a small, secondary peak at $\varpi=44~{\rm km}$, which is the 
corotation radius at which $\mbox{Re}(\omega)+2\Omega=0$ for this neutron star. 
This secondary peak is caused by the resonant response of the fluid being driven 
by the mode corotating with it (see Appendix~\ref{app:reson}).

\newpage
\begin{figure}
\epsfig{file=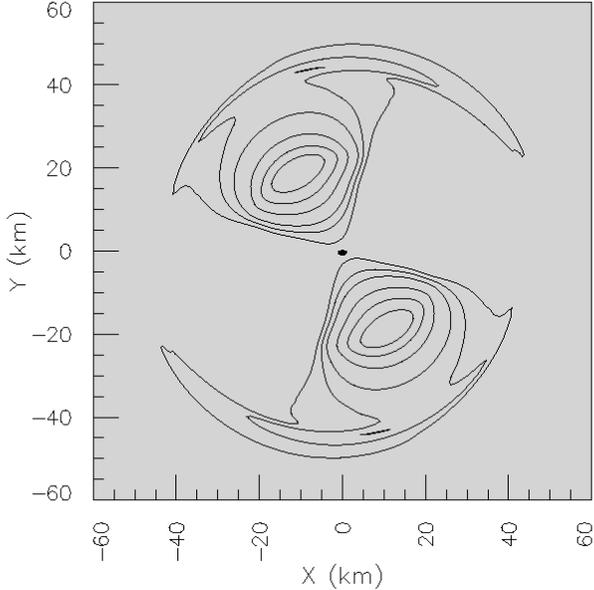,width=8cm,angle=270}
\caption{The eigenfunction of the physical density perturbation 
$\delta \tilde{\rho}$ on the equatorial plane for the bar mode 
of the most rapidly rotating star ($\beta=0.261$). The density perturbations 
are normalized so that the maximum value is one. Only positive density 
regions of the eigenfunction are shown. The negative structure of the 
eigenfunction can be inferred from the sinusoidal structure of the 
eigenfunction. The contour levels are, from inward to outward, 
0.8, 0.6, 0.4, 0.2, 0.1 and 0.01. The small arcs inside the 0.1 
contours are additional contours of 0.2, corresponding to the secondary peak 
in Fig.~\ref{fig:cdeneq}.} 
\label{fig:deneq}
\end{figure}
\vskip 0.5cm
\begin{figure}
\epsfig{file=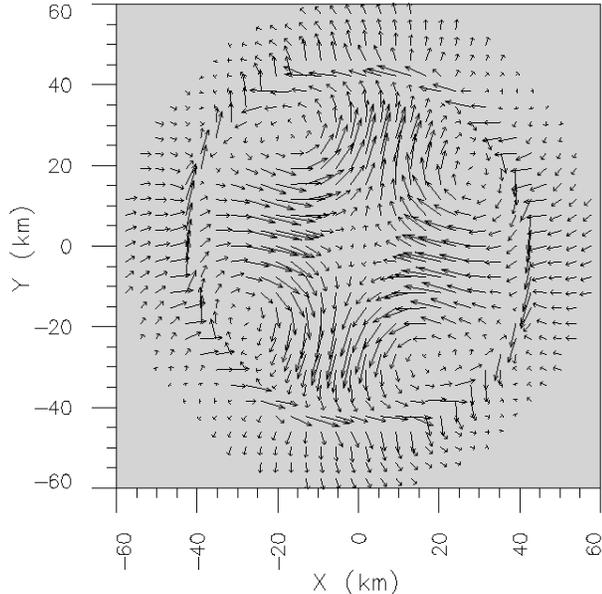,width=8cm,angle=270}
\caption{The eigenfunction of the physical velocity perturbation 
$\delta \tilde{v}^a$ on the equatorial  
plane for the bar mode of the most rapidly rotating star ($\beta=0.261$).} 
\label{fig:veleq} 
\end{figure}

Figures~\ref{fig:deneq} and~\ref{fig:veleq} show the eigenfunctions of 
the physical perturbations $\delta \tilde{\rho}$ and $\delta \tilde{v}^a$ on 
the equatorial plane. Note that our grid extends out to 200~km from the 
center, but the dynamically interesting region is concentrated within 60~km 
from the center. Since the time dependence of the perturbations go as 
$\exp[i(\omega t+m\varphi)]$, $\rm{Re}(\omega)<0$ means the pattern rotates
in prograde (counter-clockwise) direction. The density perturbation is bar-like 
in the inner 
region and becomes trailing spirals in the outer region. Similar structure 
is also observed in other numerical simulations on the bar-mode 
instability~\cite{pickett96,toman98,houser96,houser98,new00,smith96,imamura00}. 
The secondary peak of $\delta \rho$ appears as two small arcs in Fig.~\ref{fig:deneq} 
just inside the 0.1 contours. Figure~\ref{fig:veleq} shows that $\delta \tilde{v}^a$ 
is almost parallel to the $\varphi$ direction at the corotation radius,
which is also a result of resonance (see Appendix~\ref{app:reson}). Since 
$\delta \tilde{v}^a$ changes abruptly near the corotation radius, it is very 
possible that shocks will develop there when the perturbations become large. 
This might have significant influence on the non-linear evolution of the bar mode.

\begin{table}
\caption{The oscillation frequency $f$ and e-folding time $\tau$ of the most 
unstable bar mode for several unstable neutron stars. Here $\Omega$ is the 
rotational frequency of the pre-collapse white dwarf, and $\Omega_m$ is 
the maximum frequency of the white dwarf in Sequence III.}
\label{tab:omega}
\begin{tabular}{ccccc}
 $(\Omega/\Omega_m)_{WD}$ &  $\beta$ & Re($-\omega$) & $f$ & $\tau$ \\
  & & $\rm{rad}~\rm{s}^{-1}$ & Hz & ms \\
\hline
 0.934 & 0.251 & 2800 & 445 & 20 \\
 0.964 & 0.255 & 2850 & 450 & 12 \\
 0.989 & 0.258 & 2850 & 450 & 8.9 \\
 1.000 & 0.261 & 2890 & 460 & 7.8 \\
\end{tabular}
\end{table}

The eigenfunctions of the most unstable bar mode for the other unstable equilibrium 
neutron stars are similar to those displayed above. Table~\ref{tab:omega} summarizes 
the oscillation frequencies [$f=|\rm{Re}(\omega)|/(2\pi)$] and e-folding time $\tau$ 
of the unstable models we have studied. The table also shows the ratio of the rotational 
frequency of the pre-collapse white dwarfs to the maximum
frequency $\Omega_m$ of the white dwarf in the sequence. 
We find that the oscillation 
frequencies are almost the same ($\approx 450~\rm{Hz}$) for all 
the cases. We do not observe any instability in our simulations 
for stars with $\beta \leq 0.241$. Hence we conclude that $\beta_d$ is 
somewhere between 0.241 and 0.251, and the pre-collapse white dwarf has to have
$\Omega \ga 0.93 \Omega_m$ in order for the collapsed star to 
develop a dynamical instability.

\section{Gravitational Radiation}
\label{sec:gw}

In this section, we estimate the strength of the gravitational radiation 
emitted by neutron stars undergoing a dynamical instability. We also 
estimate the signal to noise ratio and discuss the detectability of 
these sources. 

The rms amplitude of a gravitational wave strain, $h(t)$, depends on the 
orientation of the source and its location on the detector's sky. When 
averaged over these angles, its value is given by~\cite{thorne87} 
\begin{equation}
  h^2(t)=\frac{1}{5} \langle h^2_+(t)+h^2_{\times}(t) \rangle \ ,
\end{equation}
where $h_+(t)$ and $h_{\times}(t)$ are the rms amplitudes of the plus and cross 
polarizations 
of the wave respectively, and $\langle ... \rangle$ denotes an average over 
the orientation of the source and its location on the detector's sky. 

In the presence of perturbations, the density and velocity of fluid inside 
the star become
\begin{eqnarray}
  \rho(\ve{x},t) &=& \rho_0(\varpi,z)+\sum_{m=-\infty}^{\infty} \delta \rho_m(\ve{x},t)
 \label{eq:rho} \\
  \ve{v}(\ve{x},t) &=& \varpi \Omega(\varpi) \ve{e_{\hat{\varphi}}}+
\sum_{m=-\infty}^{\infty} \delta \ve{v}_m(\ve{x},t) \ ,
\end{eqnarray}
where the perturbation functions $\delta \rho_m$ and $\delta \ve{v}_m$ have 
angular dependence $e^{im\varphi}$. The amplitude of the gravitational waves 
produced by time varying mass and current multipole moments can be 
derived from Ref.~\cite{thorne80}. The result is
\begin{equation}
  h^2=\frac{1}{D^2}\sum_{l=2}^{\infty}\sum_{m=-l}^l 
\frac{4G^2}{5 c^{2l+4}}N_l \left[ |D^{(l)}_{lm}|^2+|S_{lm}^{(l)}|^2\right]
\label{eq:h0}
\end{equation}
where $D$ is the distance between the source and detector; $c$ is the speed 
of light, and
\begin{eqnarray}
  N_l &=& \frac{4\pi (l+1)(l+2)}{l(l-1) [(2l+1)!!]^2} \ ; \\
  D_{lm}^{(l)} &=& \frac{d^l}{dt^l} D_{lm} \ ; \\
  S_{lm}^{(l)} &=& \frac{d^l}{dt^l} S_{lm} \ .
\end{eqnarray}
For a Newtonian source, the mass moments $D_{lm}$ and current moments $S_{lm}$ 
are given by 
\begin{eqnarray}
  D_{lm} &=& \int \rho r^l Y_{lm}^*\, d^3 x \label{eq:dlm} \\
  S_{lm} &=& \frac{2}{c}\sqrt{\frac{l}{l+1}}\int r^l \rho \ve{v}\cdot 
\ve{Y^{B}_{lm}}^*\, d^3 x \ ,
\end{eqnarray}
where $\ve{Y^{B}_{lm}}=\ve{x}\times \ve{\nabla} Y_{lm}/\sqrt{l(l+1)}$ are
the magnetic type vector spherical harmonics. The functions $D_{lm}$ and 
$S_{lm}$ have the property that $D_{lm}^*=(-1)^m D_{l\, -m}$ and
$S_{lm}^*=(-1)^m S_{l\, -m}$. Hence it is sufficient to consider only 
positive values of $m$ and Eq.~(\ref{eq:h0}) becomes 
\begin{equation}
  h^2=\frac{1}{D^2}\sum_{l=2}^{\infty}\sum_{m=0}^l 
\frac{8G^2}{5 c^{2l+4}}N_l \left[ |D^{(l)}_{lm}|^2+|S_{lm}^{(l)}|^2\right] \ .
\label{eq:h}
\end{equation}

The energy and angular momentum carried by the gravitational 
waves can also be derived from~\cite{thorne80}. The result is 
\begin{eqnarray}
  \dot{E} &=& \sum_{l=2}^{\infty} \sum_{m=0}^l \frac{G}{c^{2l+1}}\, 2N_l 
\overline{ \left[ |D_{lm}^{(l+1)}|^2 + |S_{lm}^{(l+1)}|^2\right]} \ ;
\label{eq:Edot} \\
  \dot{J} &=& \sum_{l=2}^{\infty} \sum_{m=0}^l \frac{G}{c^{2l+1}}\, 2imN_l 
\overline{
\left[ D_{lm}^{(l)*}D_{lm}^{(l+1)}+S_{lm}^{(l)*}S_{lm}^{(l+1)}\right]} \ ,
\label{eq:Jdot0}
\end{eqnarray}
where the overline denotes time average over several periods.

When a neutron star develops a dynamical instability and the bar mode ($m=2$) 
is the only unstable mode, the values of $h$, $\dot{E}$ and $\dot{J}$ will 
be dominated by the term involving $D_{22}$. Since the unstable bar mode 
has even parity under reflection about the equatorial plane, $D_{32}=S_{22}=0$ 
and the next leading term will involve $S_{32}$ and $D_{42}$. These terms 
are expected to be smaller than the $D_{22}$ term by a factor of $(v/c)^4$ 
for $\dot{E}$ and $\dot{J}$, and a factor of $(v/c)^2$ for $h$. 
In our models, $v/c < 0.28$, so the contribution of higher order 
mass and current multipole moments are small and will be neglected. 

Strictly speaking, the above analysis only applies when the amplitudes 
of the perturbations are small. When the amplitudes are large, however, 
the fluid motion does not separate neatly into decoupled Fourier 
components, so all $D_{lm}$ and $S_{lm}$ will contribute. However, 
it is expected that the $D_{22}$ term will still be the most important term. 
Since the detailed non-linear evolution of the dynamical 
instability is not known, the aim of this section is to provide an order of 
magnitude estimate of the gravitational radiation from these sources. Hence 
we shall only consider the effect of the mass quadrupole moment and assume 
$D_{22}$ can be approximated by the bar-mode eigenfunctions computed in 
Sec.~\ref{sec:LSAresult}.
In this approximation, Eqs.~(\ref{eq:h})--(\ref{eq:Jdot0}) become 
\begin{eqnarray}
  h &=& \frac{32\pi^2 G}{5c^4 D} f^2 |D_{22}|\sqrt{\frac{\pi}{15}} \ ;
\label{eq:h2} \\
  \dot{E} &=& \frac{1024\pi^7 G}{75 c^5} f^6 |D_{22}|^2 \ ; \\
 \dot{J} &=& \frac{1024\pi^6 G}{75c^5} f^5 |D_{22}|^2 \ ,
\label{eq:Jdot}
\end{eqnarray}
where $f=|\mbox{Re}(\omega)|/(2\pi)$ is the oscillation frequency of the bar mode.

Substituting the bar-mode eigenfunctions (from Sec.~\ref{sec:LSAresult})  
into Eq.~(\ref{eq:dlm}), we find that 
\begin{equation}
  |D_{22}| \approx \alpha \, 8\times 10^{45}~\rm{g}~\rm{cm}^2
\label{res:d22}
\end{equation}
for all the unstable models we have studied. Here $\alpha$ is the amplitude of the
bar mode defined in Eq.~(\ref{def:alpha}). The mass quadrupole moment $D_{22}$ 
has a time dependence $\exp(i\omega t)$, where $\omega$ is the angular frequency 
of the mode. Hence the time derivative $D_{22}^{(l)}=(i\omega)^l D_{22}$ and 
we obtain 
\begin{eqnarray}
  h &\approx & \alpha \, 7\times 10^{-23}\left( \frac{20~\rm{Mpc}}{D}\right) \ ; \\ 
  \dot{E} &\approx & \alpha^2\, 9\times 10^{52}~\rm{erg}~\rm{s}^{-1} \ ; \\ 
  \dot{J} &\approx & \alpha^2\, 6\times 10^{49}~\rm{g}~\rm{cm}^2~\rm{s}^{-2} \ . 
\end{eqnarray}

The signal to noise ratio of these sources depends on the detailed evolution 
of the bar mode when the density perturbation reaches a large amplitude and  
non-linear effects take over. Recently, New, Centrella and Tohline~\cite{new00} 
and Brown~\cite{brown00} perform long-duration simulations of the bar-mode 
instability. They find that the mode saturates when the density perturbation 
is comparable to the equilibrium density, and the mode pattern persists, 
giving a long-lived gravitational wave signal. Here we assume that this 
is the case, and that the mode dies out only after a substantial amount of angular 
momentum is removed from the system by gravitational radiation. We then 
follow the method described in Refs.~\cite{owen98,owen01} to estimate the signal 
to noise ratio.

In the stationary phrase approximation, the gravitational wave in the frequency
domain $\tilde{h}(f)$ is related to $h(t)$ by 
\begin{equation}
h^2(t) = |\tilde{h}(f)|^2 \left| \frac{df}{dt} \right| \ .
\label{eq:stapp}
\end{equation}
Combining Eqs.~(\ref{eq:h2}), (\ref{eq:Jdot}) and (\ref{eq:stapp}), we obtain
\begin{equation}
  |\tilde{h}(f)|^2=\frac{G}{c^3} \frac{\dot{J}}{5\pi f |\dot{f}| D^2} \ .
\end{equation}
The signal to noise ratio is given by 
\begin{equation}
  \left( \frac{S}{N} \right)^2 = 2\int_{0}^{\infty} \frac{|\tilde{h}(f)|^2}{S_h(f)}
\, df \ ,
\label{def:SoN} 
\end{equation}
where $S_h(f)$ is the spectral density of the detector's noise. If we assume 
that the oscillation frequency remains constant in the entire evolution, 
we obtain~\cite{blandford}
\begin{equation}
  \frac{S}{N} = \frac{1}{D} \sqrt{\frac{2G}{5\pi c^3}\frac{\Delta J}{fS_h(f)}} \ ,
\end{equation}
where $\Delta J$ is the total amount of angular momentum emitted by gravitational
waves. To estimate $\Delta J$, we assume that the mode dies out when the 
angular momentum of the star decreases to $J_d \approx 3.3\times 
10^{49}~\rm{g}~\rm{cm}^2~\rm{s}^{-1}$, which is the angular momentum of the 
marginally bar-unstable star. Then we have 
$\Delta J \la 5\times 10^{48}~\rm{g}~\rm{cm}^2~\rm{s}^{-1}$ 
for all the unstable stars, and the signal to noise ratio for LIGO-II broad-band 
interferometers~\cite{gustafson99} is 
\begin{eqnarray}
 \frac{S}{N} &=& 15 \left( \frac{20~\rm{Mpc}}{D} \right) \left( \frac{\Delta J}{5\times
10^{48}~\rm{cgs}}\right)^{1/2} \times \cr
  & & \left( \frac{f}{450~\rm{Hz}} \right)^{-1/2}
\left( \frac{\sqrt{S_h(f)}}{2\times 10^{-24}~\rm{Hz}^{-1/2}}\right)^{-1} \ .
\end{eqnarray}

The timescale of the gravitational wave emission can be estimated by the 
equation
\begin{eqnarray}
  \tau_{GW} &\sim & \frac{\Delta J}{\dot{J}} \cr
 	& \sim & 7~\rm{s}\, \left(\frac{\alpha_s}{0.1}\right)^{-2}
\left(\frac{\Delta J}{5\times 10^{48}~\rm{cgs}}\right) \ ,
\label{eq:tauGW}
\end{eqnarray}
where $\alpha_s$ is the amplitude $\alpha$ of the density perturbation 
at which the mode saturates. We have used Eqs.~(\ref{eq:Jdot}) 
and~(\ref{res:d22}) to calculate the numerical value in the last equation.

The detectability of this type of sources also depends on the event rate. 
The event rate for the AIC in a galaxy is estimated to be between $10^{-5}$ and 
$10^{-8}$ per year~\cite{kalogera99,fryer99}. Of all the AIC events, only those 
corresponding to the collapse of rapidly rotating O-Ne-Mg white dwarfs can 
end up in the bar-mode instability, and the fraction of which is unknown.
If a signal to noise ratio of 5 is required to detect the source, an event 
rate of at least $10^{-6}$/galaxy/year is required for such a source to occur 
at a detectable distance per year. Hence 
these sources will not be promising for LIGO II if the event rate is much less 
than $10^{-6}$ per year per galaxy.

The event rate of the core collapse of massive stars is much higher than that 
of the AIC.
The structure of a pre-supernova core is very similar to that of a 
pre-collapse white dwarf, so
our results might be applicable to the neutron stars produced by the core 
collapse. If the core is rapidly rotating, the resulting neutron star might be
able to develop a bar-mode instability. If a significant fraction of
the pre-supernova cores are rapidly rotating, the chance
of detecting the gravitational radiation from the bar-mode instability
might be much higher than expected. 

\section{Magnetic Field Effects}
\label{sec:mag}

As mentioned in Sec.~\ref{sec:ns}, a new-born hot proto-neutron star 
is dynamically stable because its $\beta$ is too small. It takes about 
20~s for the proto-neutron star to cool down and evolve into a cold 
neutron star, which may have high enough $\beta$ to trigger a dynamical 
instability. The proto-neutron stars, as well as the cold neutron stars 
computed in Sec.~\ref{sec:ns}, show strong differential 
rotation (Paper~I). This differential rotation will cause a frozen-in 
magnetic field 
to wind up, creating strong toroidal fields. This process will result 
in a re-distribution of angular momentum and destroy the differential 
rotation. If the timescale of this magnetic braking is shorter than the 
cooling timescale, 
the star may not be able to develop the dynamical instability discussed 
in Secs.~\ref{sec:stab} and~\ref{sec:gw}. In this Section, we estimate 
the timescale of this magnetic braking.

In the ideal magnetohydrodynamics limit, the magnetic field lines are frozen 
into the moving fluid. The evolution of magnetic field $B$ is governed 
by the induction equation 
\begin{equation}
  \frac{\partial B^a}{\partial t}+v^b \nabla_b B^a-B^b\nabla_bv^a =-B^a \nabla_b v^b \ .
\label{eq:Bind}
\end{equation}
In our equilibrium models, $v^b=\varpi \Omega(\varpi) e^b_{\hat{\varphi}}$. 
Hence $\nabla_b v^b=0$ and Eq.~(\ref{eq:Bind}) becomes 
\begin{equation}
  \frac{d B^a}{dt}=B^b \nabla_b v^a \ ,
\label{eq:Bind2}
\end{equation}
where $d/dt=\partial / \partial t + v^b \nabla_b$ is the time derivative 
in the fluid's co-moving frame. Eq.~(\ref{eq:Bind2}) can be integrated 
analytically (see e.g.\ Appendix~B of~\cite{rezzolla01}). The magnetic 
field $B^j(\ve{x},t)$ at the position $\ve{x}$ of a fluid element at time 
$t$ is related to the magnetic field $B^k(\ve{x_0},t_0)$ at the position 
$\ve{x_0}$ of the same fluid at time $t_0$ by 
\begin{equation}
  B^j(\ve{x},t)=B^k(\ve{x_0},t_0) \frac{\partial x^j}{\partial x_0^k}\ ,
\label{eq:Bevol}
\end{equation}
where $\partial x^j/\partial x_0^k$ is the coordinate strain between $t_0$ 
and $t$. With $v^a=\varpi \Omega(\varpi) e^a_{\hat{\varphi}}$, it is easy 
to show that the induced magnetic field has components only in the 
$e^a_{\hat{\varphi}}$ direction. Its magnitude $B_i$, after a time $t$, is easy 
to compute from Eq.~(\ref{eq:Bevol}). The result is
\begin{equation}
  B_i(t)= B_0 t \varpi |\partial_{\varpi}\Omega| \ ,
\label{eq:Bi}
\end{equation} 
where $B_0$ is the component of magnetic field in the $\ve{e}_{\varpi}$ direction.
The induced magnetic field will significantly change the equilibrium velocity field 
when the energy density of magnetic field $\epsilon_B=B_i^2/(8\pi)$ is comparable 
to the rotational kinetic energy density $\epsilon_R=\rho \varpi^2 \Omega^2/2$. 
This will occur in a timescale $\tau_B$ set by $\epsilon_B=\epsilon_R$. Using 
Eq.~(\ref{eq:Bi}), we obtain
\begin{equation}
  \tau_B=\frac{\Omega}{|\partial_{\varpi}\Omega|}\, \frac{\sqrt{4\pi \rho}}{B_0} 
 = \frac{L}{v_A} \ ,
\label{eq:tauB}
\end{equation}
where $L=\Omega/|\partial_{\varpi}\Omega|$ is the length scale of differential 
rotation, and $v_A=B_0/\sqrt{4\pi \rho}$ is the speed of Alfv\`{e}n waves.

Observational data suggest that the magnetic fields of most white dwarfs 
are smaller than $10^5~\rm{G}$, although a small fraction of ``magnetic white 
dwarfs'' can have fields in the range $10^6 - 10^9~\rm{G}$. Assuming flux 
conservation, the magnetic fields of the hot proto-neutron stars just 
after collapse would be $B_0 \sim 10^9~\rm{G}$ for those $10^5~\rm{G}$ white 
dwarfs. Using the angular 
velocity distribution in Paper~I for the hot proto-neutron star, 
we find that the magnetic timescale in the dynamically important region 
($\varpi\la 100~\rm{km}$) is
\begin{equation}
  \tau_B \approx 10^4~\mbox{s}\, \left(\frac{10^9~\mbox{G}}{B_0}\right) \ ,
\label{res:tauB}
\end{equation}
which is much longer than the neutrino cooling timescale ($\sim 20~\rm{s}$). 
Hence the angular momentum transport caused by the magnetic field is negligible 
during the cooling period. The magnetic timescale for the cold neutron stars 
can be calculated from the angular frequency distribution computed in 
Sec.~\ref{sec:ns}. We find that $\tau_B$ for the cold models is about 
half of that given by Eq.~(\ref{res:tauB}), which is still much longer than 
the timescale of gravitational waves $\tau_{GW}$ calculated in the previous 
Section. The instability results presented in the previous two Sections remain 
unchanged unless the neutron star's initial 
magnetic field $B_0$ is greater than $10^{12}~\rm{G}$. In that case, 
a detailed magnetohydrodynamical simulation has to be carried out to 
compute the angular momentum transport. 

The magnetic timescale for these nascent neutron stars is significantly
different from that estimated by Baumgarte, Shapiro and 
Shibata~\cite{baumgarte00} and Shapiro~\cite{shapiro01}. They consider 
differentially rotating ``hypermassive'' neutron stars, which could be 
the remnants of the coalescence of binary neutron stars. 
Those neutron stars are very massive ($M \sim 3M_{\odot}$) and have much 
higher densities than the new-born neutron stars studied in this paper. 
They also use a seed magnetic field of strength $B_0 \sim 10^{12}~\rm{G}$, which 
is much larger than our estimate. These two differences combined make our magnetic 
braking timescale two orders of magnitude larger than theirs.
It should be noted that it is the 
magnetic field just after the collapse that is relevant to our analysis 
here. The strong differential rotation of the neutron star will 
eventually generate a very strong toroidal field ($B_i \sim 10^{16}~\rm{G}$)  
and destroy the differential 
rotation. The final state of the neutron star will be in rigid rotation, and 
its magnetic field will be completely different from the initial field. For 
this reason, the field strength $B\sim 10^{12}~\rm{G}$ observed in a typical 
pulsar is probably not relevant here.

\section{Summary and Discussion}
\label{sec:dis}

We have applied linear stability analysis to study the dynamical stability 
of new-born neutron stars formed by AIC. We find that a neutron star has a 
dynamically unstable bar mode if its $\beta$ is greater than the 
critical value $\beta_d\approx 0.25$. In order for the neutron star 
to have $\beta>\beta_d$, the pre-collapse white dwarf must 
be composed of oxygen, neon, magnesium and have a rotational angular 
frequency $\Omega \ga 3.4~\rm{rad}~\rm{s}^{-1}$, corresponding to 93\% 
of the maximum rotational frequency the white dwarf can have without mass 
shedding. 

The eigenfunction of the most unstable bar mode is concentrated 
within a radius $\varpi \la 60~\rm{km}$. The oscillation frequency 
of the mode is $f\approx 450~\rm{Hz}$. When the amplitude of the 
mode is small, it grows exponentially with an e-folding time 
$\tau\approx 8~\rm{ms}$ for the most rapidly rotating star 
($\beta=0.261$), which is about 5.5 rotation periods at the 
center of the star.

The signal to noise ratio of the gravitational waves emitted by 
this instability is estimated to be 15 for LIGO-II broad-band interferometers
if the source is located in the Virgo cluster of galaxies ($D=20~\rm{Mpc}$). 
The detectability of these sources also depends on the event rate. The event 
rate of AIC is between $10^{-5}$ and $10^{-8}/\rm{galaxy}/\rm{year}$. 
Only those AIC events corresponding to the collapse of rapidly rotating 
O-Ne-Mg white dwarfs can end up in the bar-mode instability. While it is 
likely that the white dwarfs would be spun up to rapidly rotation by the accretion 
gas prior to collapse~\cite{fryer01}, it is not clear how many of the AIC 
events are related to the O-Ne-Mg white dwarfs.
If the event rate is less than $10^{-6}/\rm{galaxy}/\rm{year}$, it is not 
likely that LIGO~II will detect these sources. However, the event rate of the 
core collapse of massive stars is much higher than that of the AIC. A bar-mode 
instability could develop for neutron stars formed from the collapse of rapidly 
rotating pre-supernova cores. If a significant fraction of the cores are 
rapidly rotating, the chance of detecting the gravitational radiation 
from bar-mode instability would be much higher.

If the pre-collapse white dwarf is differentially rotating, the resulting 
neutron star can have a higher value of $\beta$. The bar-mode instability 
is then expected to last for a longer time. However, any differential 
rotation will be destroyed by magnetic fields in a timescale 
$\tau_B\sim R/v_A$, where $R$ is the size of the white dwarf and 
$v_A=B/\sqrt{4\pi \bar{\rho}} \sim B\sqrt{R^3/(3M)}$. For a massive 
white dwarf with $M=1.4M_{\odot}$, 
\begin{equation}
  \tau_B \sim 2~\rm{yrs}\ \left( \frac{10^5~\rm{G}}{B}\right)
\left(\frac{R}{1500~\rm{km}}\right)^{-1/2} \ ,
\end{equation}
which is much shorter than the accretion timescale. Hence rigid rotation 
is a good approximation for pre-collapse white dwarfs.

The magnetic field of a neutron star is much stronger than that of a white dwarf. 
The timescale for a magnetic field to suppress differential rotation depends 
on the initial magnetic field $B_0$ of the proto-neutron star. If the 
magnetic field of the pre-collapse white dwarf is of order $10^5~\rm{G}$, 
the initial field will be $B_0\sim 10^9~\rm{G}$ according to conservation
of magnetic flux. In this case, the magnetic timescale is $\tau_B \sim 10^4~\rm{s}$. 
This timescale is much longer than the time required for a hot proto-neutron 
star to cool down and turn into a cold neutron star, and go through 
the whole dynamical instability phase. If $B_0\ga 10^{12}~\rm{G}$, 
a significant amount of angular momentum transport will take place 
during the cooling phase. A detailed magnetohydrodynamical simulation 
has to be carried out to study the transport process in this case. However, such a 
strong initial magnetic field is possible only if the pre-collapse white 
dwarf has a magnetic field $B\ga 10^8~\rm{G}$.

Finally, we want to point out that the collapse of white dwarfs will certainly 
produce asymmetric shocks and may eject a small portion of the mass. We expect 
that our neutron star models describe fairly well the inner cores of the stars 
but not the tenuous outer layers. Our stability results are sensitive to the 
region with $\varpi \la 100~\rm{km}$. The results could change considerably 
if the structure in this region is very different from that of our models. 
This issue will hopefully be resolved by the future full 3D AIC simulations.

\section*{Acknowledgments}

I thank Lee Lindblom for his guidance on all aspects 
of this work. I also thank Kip S.\ Thorne and Stuart L.\ Shapiro 
for useful discussions. This research was supported by NSF grants 
PHY-9796079 and PHY-0099568, and NASA grant NAG5-4093.

\appendix

\section{Resonance at the corotation radius}
\label{app:reson}

We see from Figs.~\ref{fig:cdeneq}-\ref{fig:veleq} that the bar-mode 
eigenfunction has peciliar structures at the corotation radius ($\varpi \approx 
40~{\rm km}$) at which $\omega + 2\Omega \approx 0$. The density perturbation 
has a small peak and the velocity perturbation is almost parallel to the  
$\varphi$ direction. In this Appendix, we shall show that these are 
caused by the resonance of the fluid driven by the mode.

For simplicity, we only consider the fluid's motion on the equatorial plane. 
Assume that the perturbations are dominated by a mode that goes as 
$\exp(i\omega t + im\varphi)$. We also assume that this mode is even under 
the reflection $z \rightarrow -z$. Hence we have $\xi^z=0$ and $\delta v^z=0$. 
In cylindrical coordinates, the linearized Euler equation takes the form
\begin{eqnarray}
 i(\omega+m\Omega)\delta v^{\varpi} &=& 2\Omega \delta v^{\hat{\varphi}}
-\frac{\partial_{\varpi} \delta P}{\rho} +\frac{\partial_{\varpi} P}{\rho^2}
\delta \rho - \partial_{\varpi} \delta \Phi \label{eq:euler1} \ , \\
 i(\omega+m\Omega)\delta v^{\hat{\varphi}} &=& -(\varpi \partial_{\varpi} \Omega
+2\Omega) \delta v^{\varpi} -\frac{im}{\varpi} \left( \frac{\delta P}{\rho}
+ \delta \Phi \right) \ . \label{eq:euler2}
\end{eqnarray}
The density perturbation $\delta \rho$ is related to the pressure
perturbation $\delta P$ by
\begin{equation}
  \delta \rho = \frac{\rho}{\gamma_p P}\delta P - \left(1-\frac{\gamma_{\rm eq}}
{\gamma_p}\right) \xi^{\varpi}\partial_{\varpi} \rho \label{eq:drhomode} \ .
\end{equation}
The $\varpi$-component of the Lagrangian displacement is given by
\begin{equation}
  \xi^{\varpi} = \frac{\delta v^{\varpi}}{i(\omega+m\Omega)} \ .
\label{eq:ximode}
\end{equation}

Our numerical simulations show that $\delta P$ is well-behaved and smooth 
near the corotation radius at which $\omega +m\Omega \approx 0$. The perturbed 
gravitational potential $\delta \Phi$ is expected (and is confirmed by our 
numerical simulations) to be smooth since 
it depends on the overall distribution of the density perturbation. We can 
then use Eqs.~(\ref{eq:euler1})-(\ref{eq:ximode}) to express all the other 
perturbed quantities in terms of $\delta P$ and $\delta \Phi$. Near the 
corotation radius, the expressions are: 
\begin{eqnarray}
   \delta \rho &=& \frac{\rho}{\gamma_p P}\delta P - \left(1-\frac{\gamma_{\rm eq}}
{\gamma_p}\right) \frac{\delta v^{\varpi}\partial_{\varpi} \rho}
{i(\omega +m\Omega)} \ , \\
  \delta v^{\varpi} &\approx& \frac{-2im\Omega}{\varpi(\kappa^2+B)}
\left( \frac{\delta P}{\rho}+\delta \Phi \right) \ , \\
 \delta v^{\hat{\varphi}} &=& \frac{i}{\omega+m\Omega}\left[ 
\frac{\kappa^2}{2\Omega}\delta v^{\varpi}+\frac{im}{\varpi}\left(
\frac{\delta P}{\rho}+\delta \Phi\right)\right] \label{eq:vphi} \ , \\
 \kappa^2 &=& \varpi \partial_{\varpi}\Omega^2+4\Omega^2 \ , \\
  B &=& \frac{\partial_{\varpi} P \partial_{\varpi}\rho}{\rho^2}\left(1-
\frac{\gamma_{\rm eq}}{\gamma_p} \right) \ .
\end{eqnarray}

It follows 
from Eqs.~(\ref{eq:ximode}) 
and~(\ref{eq:drhomode}) that if $|\delta v^{\varpi}|$ is not of order 
$(\omega+m\Omega)$ near 
the corotation radius, both $|\xi^{\varpi}|$ and $|\delta \rho|$ will be large. 
The large magnitude of the Lagrangian displacement is caused by the fluid 
being driven in resonance by the mode. The large displacement of the fluid 
causes $|\delta \rho|$ to be large due to the second term of 
Eq.~(\ref{eq:drhomode}). This term arises becuase of the different 
compressibilities of stationary and oscillating fluid (i.e.\ $\gamma_{\rm eq} 
\neq \gamma_p$). 
In the case of the bar mode ($m=2$), the corotation radius is located 
at $\varpi_c \approx 40~{\rm km}$. The equilibrium density on the equator 
$\rho(\varpi_c,0)\approx 10^{12}~\rm{g}~\rm{cm}^{-3}$ and the stationary 
fluid is very compressible ($\gamma_{\rm eq}\approx 0.7$). The high 
compressibility of the stationary 
fluid make the background equilibrium density $\rho$ drop rapidly as
$\varpi$ increases, i.e.\ $|\partial_{\varpi} \rho|$ is large. The oscillating 
fluid is far less compressible ($\gamma_p=1.35$). As a result, when the oscillating 
fluid moves to a new location, it does not expand or compress to an extent 
that can compensate for the difference between the background densities at the 
old and new locations. Since both $|\xi^{\varpi}|$ and 
$|\partial_{\varpi} \rho|$ are large, $\delta \rho$ is dominated by the second 
term of Eq.~(\ref{eq:drhomode}) near the corotation radius. 
This explains the narrow secondary peak of $\delta \rho$ seen in Fig.~\ref{fig:cdeneq}. 
We see from Eq.~(\ref{eq:vphi}) that $|\delta v^{\hat{\varphi}}| \gg 
|\delta v^{\varpi}|$ and 
$\delta v^{\hat{\varphi}}$ changes rapidly near the corotation radius, 
which explains the flow pattern seen in Fig.~\ref{fig:veleq}.

\end{multicols}

\end{document}